# Hybrid Boltzmann Gross-Pitaevskii Theory of Bose-Einstein Condensation and Superfluidity in Open Driven-Dissipative Systems

D. D. Solnyshkov, H. Terças, K. Dini and G. Malpuech

*Institut Pascal, CNRS and University Blaise Pascal, 63177 Aubière cedex France.*

We derive a theoretical model which describes Bose-Einstein condensation in an open driven-dissipative system. It includes external pumping of a thermal reservoir, finite life time of the condensed particles and energy relaxation. The coupling between the reservoir and the condensate is described with semi-classical Boltzmann rates. This results in a dissipative term in the Gross-Pitaevskii equation for the condensate, which is proportional to the energy of the elementary excitations of the system. We analyse the main properties of a condensate described by this hybrid Boltzmann Gross-Pitaevskii model, namely, dispersion of the elementary excitations, bogolon distribution function, first order coherence, dynamic and energetic stability, drag force created by a disorder potential. We find that the dispersion of the elementary excitations of a condensed state fulfils the Landau criterion of superfluidity. The condensate is dynamically and energetically stable as longs it moves at a velocity smaller than the speed of excitations. First order spatial coherence of the condensate is found to decay exponentially in 1D and with a power law in 2D, similarly with the case of conservative systems. The coherence lengths are found to be longer due to the finite life time of the condensate excitations. We compare these properties with the ones of a condensate described by the popular "diffusive" models in which the dissipative term is proportional to the local condensate density. In the latter, the dispersion of excitations is diffusive which as soon as the condensate is put into motion implies finite mechanical friction and can lead to an energetic instability.

## I Introduction.

Bose-Einstein condensation (BEC) and superfluidity are among the main subjects of modern condensed matter physics. Both these related phenomena are widely studied and well understood for conservative systems in the thermodynamic limit. One of the main theoretical frameworks is the well-known Gross Pitaevskii (GP) equation for the condensate wave function $\psi(r,t)$ [1]

$$i\hbar\frac{\partial\psi(r,t)}{\partial t}=-\frac{\hbar^2}{2m}\Delta\psi(r,t)+\alpha\left|\psi(r,t)\right|^2\psi(r,t)\,, \tag{1-a}$$

which provides an accurate description of the dynamics of a superfluid at the mean-field level. Here, $m$ is the particle mass and $\alpha$ is the interaction constant. Indeed, in the presence of a homogeneous propagating condensate $\psi_0 e^{i\vec{k}_0.\vec{r}}e^{-i\mu t}$, single plane-wave excitations are not eigenstates of the GP equation (contrary to the linear Schrödinger equation). The plane-wave terms $e^{i\vec{k}r}$ and $e^{i\left(2\vec{k}_0-\vec{k}\right)r}$ interfere due to the nonlinear term $\alpha\left|\psi(r,t)\right|^2\psi(r,t)$. As a result, the dispersion of the true elementary excitations (bogolons), containing both complex exponents, is linear at low *k*, verifying the Landau criterion of superfluidity. The state $\psi_0 e^{ik_0 r}e^{-i\mu t}$ is thus protected against excitations despite not being the ground state of the system. Nevertheless, there is no satisfactory model developed so far to fully describe condensation in an open-dissipative system. This is especially important, since the standard test-bed systems of superfluidity, such as liquid Helium and atomic BECs [2], or exciton-polariton BECs [3,4], are only approximately conservative.

A fundamental aspect of BECs and superfluids is the presence of a normal (non-condensed) fraction. This aspect has been first considered in Landau's two-fluid hydrodynamic theory for liquid $^4$He [5]. The two-fluid model is only valid when the collisions between the excitations occur so often that they lead to thermalization within the gas of the excitations. The evaporation of helium is simply neglected in this model, and the conservation of the total number of particles is thus required. In 1958, Pitaevskii introduced a phenomenological damping term in Eq. (1) [6] for the superfluid fraction, which drives the system towards a stationary state. This damping accounts for the decay of the excitations of the superfluid fraction due to various thermalization mechanisms. The modified GP equation is obtained by transformation of the Hamiltonian as $H^{'}=\left(1-i\Lambda\right)H$, where $\Lambda$ is a small dimensionless parameter controlling the strength of dissipation. Pitaevskii used the above mentioned Hamiltonian in the hydrodynamic limit, where the modification of the local density associated with bogolons is averaged out, since their density is very high. The dissipation term in this simplest approximation is therefore linearly proportional to the deviation of the energy from the equilibrium value (which is the chemical potential). This dissipative model has later been used to describe relaxation processes in atomic condensates [7], outside of the limits of the hydrodynamic approximation.

One of the most representative finite temperature particle-conserving theories developed later for atomic BECs is the kinetic theory by Zaremba, Nikuni and Griffin (ZNG) [8,9] based on the equation for the condensate $\psi(r,t)$ within the time-dependent Hartree-Fock-Popov approximation, where the dynamics of the condensate wave function is coupled to the thermal gas. The latter is described in terms of a semi-classical Boltzmann equation for the distribution function. Two-fluid hydrodynamic theories are successful in describing some finite-temperature features of atomic BECs [10], as it is the case of the second-sound mode [11,12], corresponding to an entropy wave between the condensed and the thermal densities.

A proper description of BEC in an open dissipative system, where the particle density results from a balance between pumping and decay mechanisms, such as polariton BECs [13,4], requires to include the two above-mentioned ingredients: the relaxation and dissipation processes for both condensed and normal parts. Exciton-polaritons are composite bosonic quasi-particles appearing in quantum microcavities in the regime of strong coupling between the excitons (quantum well or bulk ones) and cavity photons. The photon lifetime is relatively short (30 ps in best cavities), although sufficiently long for the establishment of thermal equilibrium and polariton condensation even up to the room temperature, as a consequence of the very small effective mass of polaritons ($5\times10^{-5}$ of the free electron mass). Still, a constant pumping is required in order to maintain stationary particle density in the system. This pumping is usually optical and non-resonant, which means that the pump laser creates high-energy excitons, which are not coupled to the light. These excitons thermalize very rapidly and form a reservoir, whose presence is often very important in the studies of polariton condensates. The main mechanism of the particle transfer from the reservoir to the condensate is the stimulated polariton-polariton scattering.

Fundamental properties of such pumped/decaying condensates, such as their superfluidity [14,15,16,17,18,19] or first order spatial coherence [20, 21], are indeed still a matter of debate. Aside from the useful but limited approaches based on the Boltzmann equations [22,23,24,25,26,27,28,29], most of the efforts have been put on models where an equation describing a pumped reservoir is coupled to a GP-like equation [16,17]. These models, which we will refer to as “diffusive” models, became popular and has been developed and used under several versions [18, 30, 31, 32, 33]. Within these models, the condensate dynamics can be mostly described in the two different forms:

$$i\hbar\frac{\partial\psi}{\partial t}=-\frac{\hbar^2}{2m}\Delta\psi+\alpha|\psi|^2\psi-i\Lambda\alpha\left(|\psi|^2-n_c\right)\psi \quad \text{(1-b)}$$

$$i\hbar\frac{\partial\psi}{\partial t}=\left(1-i\Lambda\right)\left(-\frac{\hbar^2}{2m}\Delta\psi+\alpha|\psi|^2\psi-\mu\psi\right) \quad . \quad \text{(1-c)}$$

Equivalent forms of 1-b has been introduced in [16,17] and will be referred to as the diffusive Ginzburg-Landau (GL) model. In such a case, the dissipation is proportional to the non-linear term of the GP equation only. (1-c) is the equivalent to the introduction of a dissipative term by Pitaevskii [6] but used beyond the hydrodynamic approximation [7, 18]. Damping is proportional to the total local energy. A recent analysis shows that, within the Bogoliubov approximation, the first order coherence extracted out of these models shows the same long range behaviour as in the equilibrium models [32]. Namely, a power-law and an exponential decay of coherence are found in 2D and 1D systems, respectively. Analysis going beyond the Bogoliubov approximation suggest that the long range decay in 2D systems is not always realized, depending on the exact system parameters [33]. These models provide a diffusive dispersion for the elementary excitations of a condensate at rest, which does not fulfil the Landau criterion of superfluidity. As a result, disorder and defects provoke a finite dissipation of a flow (friction), even for velocities below the condensate sound speed. The use of these models therefore lead to the conclusion that a condensate in an open dissipative system must not be superfluid according to Landau's criterion, even if their wave function exhibit long-range order. Another approach to the energy relaxation has been proposed in [34], where both spontaneous and stimulated scattering with phonons were taken into account by the introduction of a stochastic variable into the GP equation, but the resulting dynamical properties of the fluid have not been analyzed.

This paper is organized as follows, In Section II, we microscopically derive a modified version of the dissipative GP equation accounting for relaxation, starting from semi-classical kinematic model of the reservoir dynamics. We refer to the latter as the hybrid Boltzmann-Gross-Pitaevskii (hybrid BGP) model throughout this work. The system of two coupled equation can be solved iteratively. At the first iteration, the critical condensation conditions are obtained. The dissipation term which is obtained is proportional to the kinetic energy only, justifying the phenomenological model we introduced in [35] to fit time-resolved data. At the second iteration, the scattering rates are recalculated according to the renormalized elementary excitation energies. The condensation conditions are not modified, but the dissipation term becomes linearly proportional to the energy of the condensate excitations (bogolons). In both cases, the dissipation term - arising from the Boltzmann scattering rates and finite particle lifetime - is not affecting the nonlinear interference term of the GP equation. The dispersion of the elementary excitations is consequently compatible with the Landau criterion of superfluidity. In the Section III, we briefly remind the main basic features of the diffusive models, computing the dispersion of the elementary excitations of the condensate at rest. In Section IV, we calculate the bogolon distribution function created by thermal excitations and induced by disorder, for both the hybrid BGP and the diffusive models. The first-order spatial coherence is computed for 1- and 2-dimensional systems. The coherence found for the three models is qualitatively similar to that of a conservative condensate. We consider the stability of the condensate when it is set into motion in Section V. Within the hybrid BGP model, the condensate is stable until it reaches the sound speed $c_s = \sqrt{\alpha n_c / m}$ , where $n_c$ is the condensate density. It becomes energetically and dynamically unstable when the flow velocity is larger than $c_s$, in contrast to the conservative case in which a condensate is only energetically unstable above the speed of sound. Condensates described by the diffusive GL model are found to be energetically unstable regardless the velocity of the flow, which is not the case for the diffusive Pitaevskii model. In Section VI, we calculate the friction induced by a disorder potential on a propagating condensate. The results of the hybrid BGP model are similar to those of conservative systems (superfluidity below $c_s$). The diffusive GL model yields a finite friction even at low speeds, but with a clear threshold behavior at $c_s$. Finally, Section VII is devoted to the summary of the results. A discussion of the implications of the present work and some conclusion remarks are stated in Section VIII.

## II Hybrid BGP Model

In what follows, we describe the small deviations from the stationary regime of a weakly-interacting bosonic gas with a finite lifetime. A constant, homogeneous reservoir is created via non-resonant pumping, providing an incoherent injection of particles. An analogue system could be a porous vessel containing liquid $^4$He (below the $\lambda$-point), through which the superfluid fraction can escape, representing losses. Pumping then corresponds to continuously refilling the vessel with helium in the normal phase.

We first consider an infinitely large homogeneous microcavity [13], in which the strong coupling regime takes place between quantum well excitons and confined cavity photons. The exciton reservoir density $n_R$ is assumed to be constant, independent of the small-amplitude fluctuations within the condensate. This reservoir is coupled to a Gross-Pitaevskii equation via the semi-classical Boltzmann rates. At a first iteration, we solve the set of two coupled equations in the condensation regime, finding the eigenstates of the system (bogolon energy and decay time) assuming that the scattering rates are not affected renormalized due the presence of the condensate. We then use the bogolon energies to calculate the scattering rates which allows to renormalize the decay rate of the bogolon states.

The homogeneous, thermalized reservoir description is a central approximation of our model, which is the limit opposite to the one used in the previous models (diffusive GL and Pitaevskii). This assumption relies on the rapid relaxation (thermalization) times within the reservoir, and its validity will be qualitatively discussed at the end the paper. We notice that such an approximation is not accurate for liquid helium, since both normal and superfluid fractions are composed of the same particles. In exciton-polaritons, however, the reservoir is composed of pure excitons, which exhibit fast thermalization [36] and fast momentum dephasing times with acoustic phonons [37], while the condensate - formed at the bottom of the polariton branch - is strongly decoupled from phonons because of the steepness of the polariton dispersion. Moreover, since the excitations of the condensate created by thermal and quantum fluctuations are not coherent, they cannot create correlated fluctuations in the reservoir, leading to the modulations of its density. In fact, polaritons propagating for hundreds of µms have been observed without any relaxation induced by phonons [21,35]. A scheme of the configuration we consider is shown in Fig. 1.

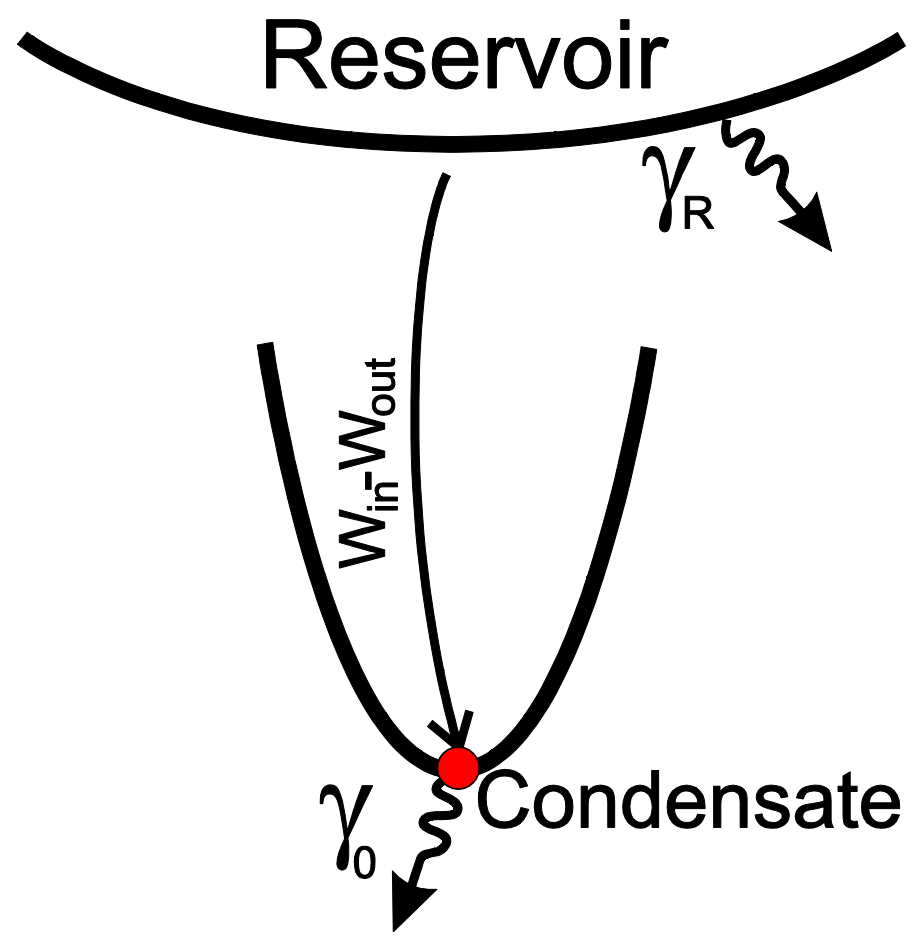

Fig. 1. Sketch of the polariton condensate with a non-resonantly pumped reservoir. The reservoir excitons, created after thermalization with the lattice phonons, decay at a rate $\gamma_R$ and scatter in (out) the condensate at a rate $W_{in}$ ($W_{out}$). The polaritons forming the condensate at the bottom of the lower polariton branch decay at a rate $\gamma_0$.

Within this approximation, the expression for the reservoir dynamics is given by

$$\frac{dn_R}{dt} = P_R - \gamma_R n_R - \int \left( W_{in}^k - W_{out}^k \right) n_c(k,t) \rho_k^d dk \ , \tag{2}$$

where the scattering rates into and out the *k*-state read $W_{in}^k = \int \tilde{W}_{in}^{k'k} n_R\left(k',t\right) \rho_{k'}^{2D} dk'$, and $W_{out}^k = \int \tilde{W}_{out}^{k'k} \left( n_R\left(k',t\right) + 1 \right) \rho_{k'}^{2D} dk'$. Here, $P_R$ is the pumping rate, $\gamma_R$ is the mean reservoir decay, $n_R\left(\vec{k},t\right)$ and $n_c\left(\vec{k},t\right)$ are the distribution function of excitons in the reservoir and of polaritons in the condensate, respectively; $\rho_k^d$ is the polariton density-of-states in a system of dimensionality *d*. $\tilde{W}_{in}^{k'k}$ and $\tilde{W}_{out}^{k'k}$ are standard semi-classical Boltzmann rates [38]. Notice that Eq. (2) can be easily applied to systems other than polaritons, provided the scattering rates are properly modified. Spontaneous scattering into polariton states is neglected at that stage. The reservoir is assumed to be thermalized at the lattice temperature *T*, so that $n_R\left(\vec{k},t\right) \approx n_R(t) \exp\left(-E_R(k)/k_B T\right)$. In the present model, we include exciton-phonon and exciton-exciton scattering processes only. Within the thermal reservoir approximation, the scattering rates are given by [39]

$$W_{in}^k \approx n_R^2 M_{XX} + n_R M_{XP} \ , \tag{3}$$

$$W_{out}^k \approx \left( n_R M_{XX} + M_{XP} \right) \frac{m_X k_B T}{2\pi\hbar^2} e^{\left(E(k) - E_X\right)/k_B T} \ , \tag{4}$$

where $M_{XX}$ and $M_{XP}$ are the exciton-exciton and exciton-phonon effective scattering rates respectively. Here, $m_X$ is the exciton mass, $E(k)$ and $E_X$ are the polariton and bare exciton energies. Equation (4) is obtained assuming that the occupancy of reservoir states is much smaller than the one of the condensate.

The dynamics of the condensate is described by the Gross-Pitaevskii equation coupled to the reservoir

$$\begin{aligned} i\hbar \frac{\partial \psi\left(\vec{k},t\right)}{\partial t} = -i\frac{\hbar\gamma_k}{2}\psi(\vec{k},t) + E\left(\vec{k}\right)\psi(\vec{k},t) + \alpha \int \left( \left|\psi(\vec{r},t)\right|^2 + n_R \right) \psi(\vec{r},t) \exp(i\vec{k}\vec{r}) d\vec{r} + \\ + i\hbar \frac{1}{2} \left( W_{in}^k - W_{out}^k \right) \psi(\vec{k},t), \end{aligned} \tag{5}$$

where $\psi(\vec{k},t) = \int \psi(\vec{r},t) \exp(i\vec{k}.\vec{r}) \ d\vec{r} \ \psi(\vec{k},t) = \int \psi(\vec{r},t) \exp(i\vec{k}.\vec{r}) \ d\vec{r}$, $\gamma_k$ is the decay rate of a polariton state *k*, $\alpha$ is the constant of the polariton-polariton interaction.

### *1. First iteration*

Our goal is to obtain a single damped Gross-Pitaevskii equation from the system of equations for the reservoir and the condensate solved for stationary situation. At the first iteration, we will neglect the renormalization of the scattering rates induced by the blueshift of the polariton energy as a consequence of condensation. The essential point is that the dissipative part of the right-hand side of the Eq. (6) depends only on the scattering rates, which by nature do not include spatial interference of the nonlinear term.

A condensed mode can be formed at a given wavevector $k$ if the positive gain condition, $\left(W_{in}^{k}-W_{out}^{k}\right)-\gamma_k \geq 0$ is fulfilled. Since the polariton decay time is mostly determined by its photonic fraction, its decay rate (at zero exciton-photon detuning) can be approximately given by [38]

$$\gamma_k \approx \gamma_0\left(1-\frac{\hbar k^2}{2m\Omega}\right), \tag{6}$$

where $m$ is the polariton mass, $\hbar\Omega$ the Rabi splitting, $\gamma_0=\gamma_c/2$ and $\gamma_c$ is the photonic decay rate of the cavity. We see that the lifetime is longer for higher wavevectors because the photonic fraction decreases. A similar $k^2$ dependence can be found (with a different coefficient) for a non-zero detuning case. In the following, we neglect the phonon scattering within the polariton branch because of the steepness of its dispersion, as discussed above.

The key approximation is the linearization of the scattering rates close to the ground state. Indeed, $E(k)/k_BT<<1$ near $k$ = 0, where the condensate is formed, and the exponent in the scattering rates is replaced by a linear dependence. Therefore, the damping of the state $k$ becomes linearly proportional to its energy, as in the original proposal of Pitaevskii [6]. Therefore, the condensate gain condition can be expressed as

$$\left(W_{in}^{k}-W_{out}^{k}\right)-\gamma_k=\kappa-\frac{\beta}{\hbar}\frac{\hbar^2k^2}{2m} \tag{7}$$

where $\kappa=n_RM_{XX}\left(n_R-f(T)\right)-\gamma_0$, $\beta=\dfrac{\hbar n_RM_{XX}f(T)}{k_BT}-\dfrac{\gamma_0}{\Omega}$, $f(T)=\dfrac{m_Xk_BT}{2\pi\hbar^2}e^{-E_X/k_BT}$.

Putting Eqs. (5) and (7) together, and performing a Fourier transform, we are finally able to write a modified GP equation for the condensate the presence of the lifetime and a homogeneous reservoir

$$\frac{dn_R}{dt}=P_R-\gamma_Rn_R-\int\left(\kappa-\beta\frac{\hbar k^2}{2m}\right)n_c(k,t)\rho_k^d dk \tag{8}$$

$$i\hbar\frac{\partial\psi}{\partial t}=-\left(1-i\beta/2\right)\frac{\hbar^2}{2m}\Delta\psi+\alpha\left(|\psi|^2+n_R+i\hbar\kappa/2\right)\psi\,. \tag{9}$$

Condensation in the ground state requires $\kappa=0,\beta\geq 0$, which yields the following simultaneous conditions for the reservoir density

$$n_R=\frac{f(T)+\sqrt{f^2(T)+4\gamma_0/M_{XX}}}{2} \quad \text{and} \quad n_R\geq\frac{\gamma_0k_BT}{\hbar\Omega M_{XX}f(T)}\,. \tag{10}$$

If $\beta\geq 0$, the system is in the situation where the gain decreases versus $k$. The ground state is more populated than the excited states, as physically expected for condensation. This situation, favored by high temperatures and large Rabi splitting Ω, is achieved if the effective scattering rate $M_{XX}$ and density $n_R$ are large enough compared with the decay rate $\gamma_0$. If $\beta < 0$, the gain is larger for excited states, and thus the bottleneck effect, leading to out-of-equilibrium condensation, is expected to occur. In the following, we are going to treat the case of the condensation in the ground state, so that the conditions of Eq. (8) are verified. This is the case for temperatures above $T_{c_1}$, with $T_{c_1}$ being the solution of the following condition

$$\hbar\Omega M_{XX} f(T_{c_1}) \frac{f(T_{c_1}) + \sqrt{f^2(T_{c_1}) + 4\gamma_0 / M_{XX}}}{2\gamma_0 k_B T_{c_1}} = 1 \tag{11}$$

The solution to Eq. (11) allows us to find the temperature range in which ground state condensation can take place versus the system parameters. Below $T_{c_1}$, the system is in the kinetic regime. The relaxation is not fast enough compared with the particle lifetime. The condensation threshold is governed by the relaxation kinetics and the critical reservoir density decreases versus temperature [40,41,42,43]. Above $T_{c_1}$, the system enters the thermodynamic regime. Ground-state condensation occurs and the reservoir density at threshold increases versus temperature [40,41,42,43]. In such a case, one can find a homogeneous solution to Eqs. (8, 9) which is obtained by taking:

$$\begin{cases} \psi(\vec{r},t) = \psi_c \exp(-i\mu t) \\ n_R = \dfrac{f(T) + \sqrt{f^2(T) + 4\gamma_0 / M_{XX}}}{2} \\ n_c = |\psi_c|^2 = \dfrac{P_R - \gamma_R n_R}{\gamma_0} \\ \mu = \alpha(n_c + n_R) \end{cases} \tag{12}$$

Fig. 2 depicts $T_{c_1}$ as given by Eq. (11) for a typical GaAs-based cavity and the temperature dependence of the reservoir density at threshold as given by (12), which depends on the pumping as $n_R = P_R / \gamma_R$. When $n_R$ becomes larger than the critical value implying the breakdown of strong coupling, our model is not applicable and one should consider electron-hole plasma physics [44]. This critical value (typically 10-50 % of the Mott density of carriers per quantum well) fixes a second critical temperature $T_{c_2}$. Therefore, the quasi-condensation in the thermodynamic regime arises only in a finite temperature range $T_{c_1} < T < T_{c_2}$. In this configuration, the balance between gain and losses is achieved only for the ground state, while the excited states have a net decay rate growing as $k^2$, since the scattering rate towards the polariton states (as a function of their energy) decreases faster than the radiative losses, as can be seen from Eq. (8). The damping we obtain is therefore linearly proportional to the energy of the excited states, which, without renormalization, grows as $k^2$. The specific $k^2$ dependence of the damping parameter, found here microscopically, justifies the phenomenological model introduced in Ref. [35].

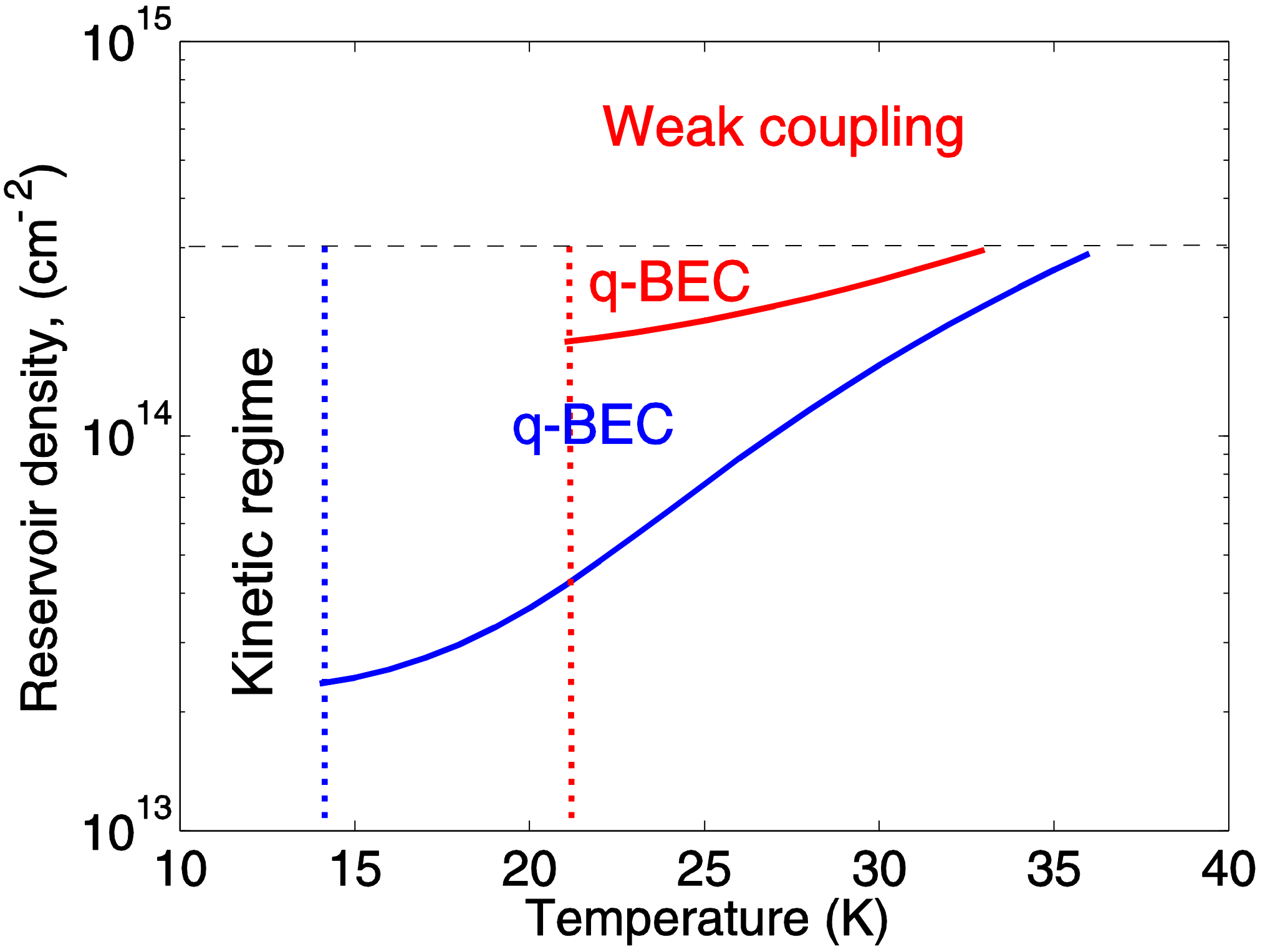

Fig. 2. (color online) Quasi-condensation phase diagram of a GaAs based microcavity. The vertical dotted lines show the critical temperature $T_{c_1}$ obtained from Eq. (11). The horizontal dashed line shows the density above which strong coupling is expected to be lost (transition towards the electron-hole plasma regime). The solid line show the critical density for quasi-condensation. We have considered the following parameters: $\hbar\Omega$=15 meV, $m_X$ = 0.5 $m_0$, $m_0$ being the free electron mass, $M_{XX}$ = $10^{-17}$ $m^4.s^{-1}$, $\gamma_0 = (100\,ps)^{-1}$ (blue/bottom solid line), $\gamma_0 = (2\,ps)^{-1}$ (red/top solid line).

In a previous work [45], a slightly different gain condition was derived, involving an excess temperature of the polariton gas with respect to the reservoir. Here, this condition is relaxed since we account for the true energy gap between the exciton reservoir and the condensate, as illustrated in Fig. 1. Nevertheless, one should note that out-of-equilibrium condensation can occur if the condition (11) is not fulfilled (for example if $\kappa \geq 0, \beta \geq 0$). In such case, complicated dynamical effects can take place (bottleneck is formed and then particles are transferred into the ground state) and can invalidate the conclusions of the present work.

### 2. *Second iteration: renormalization of the scattering rates*

Equations (8,9) have been obtained neglecting the renormalization of the energy of the excited states by the interactions within the condensate. At the second iteration, we have to include the interactions (leading to the Bogoliubov dispersion) into consideration, and introduce the decay term linearly proportional to the energies of the new elementary excitations of the system.

The dynamical equation describing the condensate evolution around the stationary homogeneous solution reads

$$i\hbar\frac{\partial\psi}{\partial t}=-(1-i\beta)\frac{\hbar^2}{2m}\Delta\psi+\alpha\left(|\psi|^2+n_R\right)\psi \quad . \qquad (13)$$

We now apply the standard linearization procedure in order to find the dispersion of the elementary excitations of the condensate. We make use of the Bogoliubov prescription,

$$\psi(\vec{r},t)=\psi_c\exp(-i\mu t)\left(1+A_{\vec{k}}\exp\left(i(\vec{k}\vec{r}-\omega_k t)\right)+B^*_{\vec{k}}\exp\left(-i(\vec{k}\vec{r}-\omega^*_k t)\right)\right) \qquad (14)$$

and keep only the first order terms in $A_k$ and $B_k$, such that $n(\vec{k}=0,t)=n_c=|\psi_c|^2$. The dispersion of the elementary excitations of the condensate then reads

$$\hbar\omega_k=-i\beta\frac{\hbar^2k^2}{2m}+\sqrt{\left(\frac{\hbar^2k^2}{2m}\right)^2+2\alpha n_c\left(\frac{\hbar^2k^2}{2m}\right)} \quad . \qquad (15)$$

For small *k*, one obtains $\hbar\omega_k\approx\pm\hbar c_s k-i\beta\frac{\hbar^2k^2}{2m}$. This is a damped Bogoliubov mode, which verifies the Landau criterion of superfluidity. Here, the speed of sound is given in terms of the condensate density only, and does not contain the effective chemical potential (corresponding to the total blue shift measured in experiments).

Once the energies of the eigenmodes are found, we can use them to re-calculate the scattering rates by replacing the real part of the polariton energy $E(k)$ in Eq. (4) by the real part of the bogolon energy in Eq. (15) [46]. Therefore, at the second iteration, the damped Gross-Pitaevskii equation takes the following shape (written in the reciprocal space):

$$i\hbar\frac{\partial\psi_k}{\partial t}=\frac{\hbar^2k^2}{2m}\psi_k+\alpha\int\left(|\psi|^2+n_R\right)\psi e^{i\vec{k}\vec{r}}\,d\vec{r}-\frac{i\beta}{2}\sqrt{\left(\frac{\hbar^2k^2}{2m}\right)^2+2\alpha n_c\frac{\hbar^2k^2}{2m}}\psi_k \, . \qquad (16)$$

As a result, the energy of the elementary excitations reads

$$\hbar\omega_k=(1-i\beta)\sqrt{\left(\frac{\hbar^2k^2}{2m}\right)^2+2\alpha n_c\left(\frac{\hbar^2k^2}{2m}\right)} \, . \qquad (17)$$

Since the real part of the bogolon energy is not affected by the renormalization, we can conclude that the two-step iterative process is convergent. Fig. 3-a shows the real and imaginary part of the spectrum (17).

The coefficients $|A_{\vec{k}}|^2, |B_{-\vec{k}}|^2$ can be found following the same procedure as in the equilibrium case, taking the normalization condition $|A_k|^2-|B_{-k}|^2=1$, which gives the same result as in the equilibrium case

$$\left|A_k\right|^2 = \frac{\left(\frac{\hbar^2 k^2}{2m} + \mathrm{Re}(\hbar\omega_{k,\pm}) + \alpha n_c\right)^2}{\left(\frac{\hbar^2 k^2}{2m} + \mathrm{Re}(\hbar\omega_{k,\pm}) + \alpha n_c\right)^2 - \left(\alpha n_c\right)^2} \approx 1 + \frac{\alpha n_c}{2\,\mathrm{Re}(\hbar\omega_{k,\pm})}, \tag{18}$$

$$\left|B_{-k}\right|^2 = \frac{\left(\alpha n_c\right)^2}{\left(\frac{\hbar^2 k^2}{2m} + \mathrm{Re}(\hbar\omega_{k,\pm}) + \alpha n_c\right)^2 - \left(\alpha n_c\right)^2} \approx \frac{\alpha n_c}{2\,\mathrm{Re}(\hbar\omega_{k,\pm})}, \tag{19}$$

where the approximate expressions are given near $k \to 0$. It is important to note that the spatial variations of the condensate density due to the presence of the bogolon do not lead to spatial variations of the energy of this bogolon. The latter is a collective state and its energy is constant in space. Therefore, the new dissipation term does not prevent superfluidity, because it does not destroy the interference term $\left|\psi\right|^2 \psi$. This is why it is possible to use the real part of the bogolon energy to renormalize the scattering rates, which in turn affects the imaginary part of the dispersion.

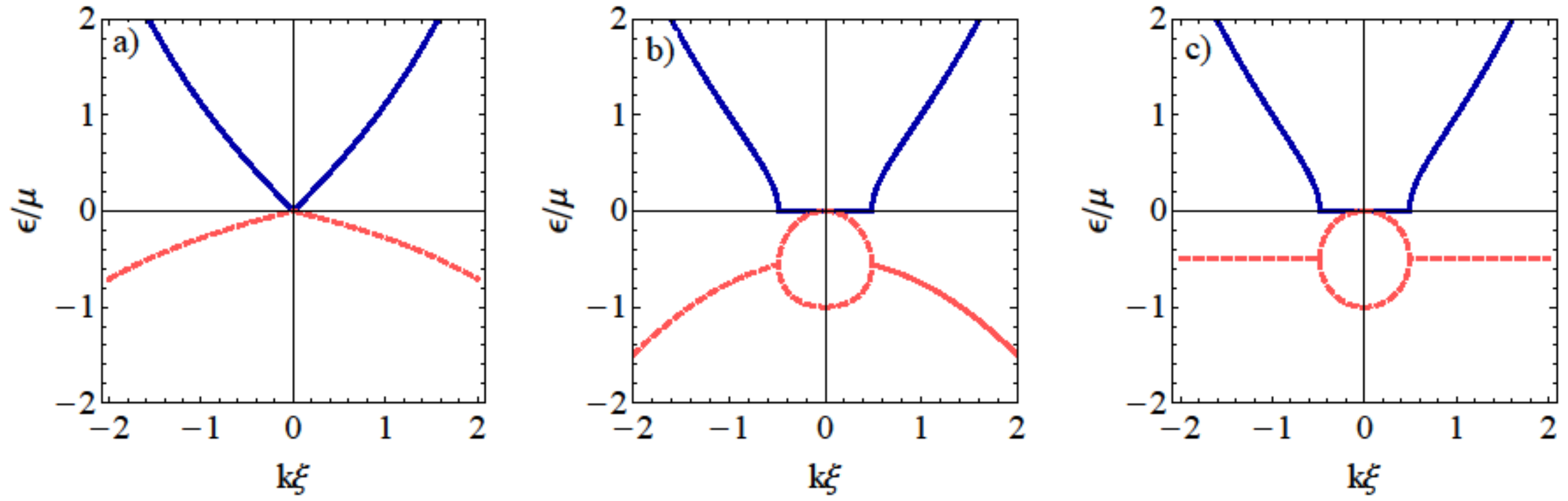

Fig. 3. (color online) Bogoliubov excitation spectra for the three different models considered in the paper. a) Hybrid Boltzmann Gross-Pitaevskii model with $\beta = 0.2$, b) diffusive Pitaevskii model of Eq. (1-c) with $\Lambda = 0.5$, and c) diffusive Ginzburg-Landau model of Eq. (1-b) with $\Lambda = 0.5$. For all the cases, the blue/solid line (red/dashed line) depicts the real (imaginary) part of the spectrum.

## III. Diffusive models for the condensate at rest

Let us compare the results of the above procedure with the damped Gross-Pitaevskii equations [6,7,18,31,32,33]. As it was mentioned in the introduction, these equations can be classified into two families, based on whether the dissipation affects the total local energy, or the local density. Both approaches provide sensibly the same qualitative properties of the system when the condensate is at rest. If the same linearization approach is applied to Eq. (1-b) or to Eq. (1-c), one finds that the dispersion is purely imaginary at small wavevectors, and the real part of the frequency is flat, which corresponds to diffusive behaviour: the density waves decay without propagation. In this section, we consider the diffusive Pitaevskii model described by Eq. (1-c), for definiteness. The dispersion of the elementary excitations is given by the following expression

$$\hbar\omega_k = -i\Lambda\left(\alpha n_c + \frac{\hbar^2 k^2}{2m}\right) \pm \sqrt{-\alpha^2 n_c^2 \Lambda^2 + \alpha n_c \frac{\hbar^2 k^2}{m} + \frac{\hbar^4 k^4}{4m^2}} \quad . \tag{20}$$

The real and imaginary parts of (21) are illustrated in Fig. 3-b), whereas the bogolon energies of the dissipative GL model are shown in Fig 3-c). The boundary between the diffusive and propagative parts of the dispersion is determined by a critical wavevector $k_c$ given by the relation (assuming that $\Lambda$ is small) $\frac{\hbar^2 k_c^2}{2m} = \alpha n_c \frac{\Lambda^2}{2}$. Such a critical wavevector can be found for arbitrarily small values of $\Lambda$, and this diffusive behaviour can be observed in numerical experiments. According to the Landau criterion, the flat part of the dispersion prevents the existence of superfluidity. This makes the model not compatible with the system for which it was initially conceived: liquid helium below the Lambda-point, where it is experimentally found to be superfluid. Such contradiction appears because the damping term in Eq. (1-c) contains a real-space dependence (see also Eq. (21)). Because of this, the damping term is not anymore proportional to the energy of the state concerned, as initially proposed by Pitaevskii. Indeed, at low wavevectors the interaction energy dominates over the kinetic energy and the spatial variations of the density of the condensate due to a bogolon imply that the energy damping term in the rhs of the Eq. (1-c) appears to be different at different points, which leads to a local dissipation instead of propagation (overdamping of the excitations). In order to use a phenomenological damping term proportional to the energy, as derived from the condition of an equilibrium reservoir, one has to use Eq. (16). In fact, the breakdown of superfluidity is due the extension of Pitaevskii's model beyond the hydrodynamic approximation.

## IV. Spatial coherence induced by thermal fluctuations and disorder

In order to estimate the decay of the first order spatial coherence, we need to evaluate the distribution function of the elementary excitations. As seen from Eqs. (16) and (17), the energy is complex, which means that the excitations on top of the condensate decay in time. The bogolon distribution function is - in general - not thermal, and rather depends on the competition between scattering rates and lifetime. The bogolon distribution function obeys the following semi-classical Boltzmann equation

$$\frac{dN_b(k)}{dt} = \mathrm{Im}(\omega_k)N_b(k) + \left(W_{b,k}^{in} - W_{b,k}^{out}\right)N_b(k) + W_{b,k}^{in} \, . \tag{21}$$

The exchange of particles with the reservoir and the lifetime are included in the term $\mathrm{Im}(\omega_k)$. The exchange of particles with the condensate is described by the rates $W_{b,k}^{in}$, $W_{b,k}^{out}$ which can have different origins. Eq. (21) admits a stationary solution if

$$W_{b,k}^{in} - W_{b,k}^{out} + \mathrm{Im}(\omega_k) \leq 0 \tag{22}$$

If the condition (22) is not verified, the condensate is unstable against the creation of bogolons. In the steady state, the distribution function of bogolons is given by

$$N_b(k) = \frac{W_{b,k}^{in}}{-\mathrm{Im}(\omega_k) - W_{b,k}^{in} + W_{b,k}^{out}} \tag{23}$$

***1. Thermal fluctuations***

In what follows, we consider the scattering rates induced by the thermal fluctuations. The thermal rates depend on the phonon scattering and on the scattering with reservoir particles. For example, a condensate particle is scattered to a bogolon state and the process is assisted by an exciton scattering process in the reservoir. As a result, for both mechanisms, we may have $\frac{W_{b,k}^{in}}{W_{b,k}^{out}} = \exp\left(\frac{-\mathrm{Re}(\hbar\omega_k)}{k_B T}\right)$. This approach is equivalent to add a thermal noise term in the GP equation [31]. The steady state bogolon distribution function therefore reads

$$N_b(k) = \frac{W_b e^{-\mathrm{Re}(\hbar\omega_k)/k_B T}}{-\mathrm{Im}(\omega_k) + W_b\left(1 - e^{-\mathrm{Re}(\hbar\omega_k)/k_B T}\right)} , \tag{24}$$

where $W_b \sim n_c n_R M_{XX}$ . Eq. (24) is a Bose distribution function characterised by a temperature *T*, but with a non-zero chemical potential $\mu_{bog} = k_B T \ln\left(1 + \frac{\beta k_B T}{W_b \hbar}\right)$ which tends to zero in the limit of infinite lifetime. The distribution (24) is quite general and can be applied to any bogolon dispersion, and also in the infinite lifetime limit regime, for which we simply recover the usual equilibrium Bose function. In the low bogolon energy limit, we may write

$$N_b(k) \approx \frac{W_b}{-\mathrm{Im}(\omega_k) + \frac{\hbar W_b}{k_B T}\mathrm{Re}(\omega_k)} . \tag{25}$$

Using the bogolon dispersion in Eq. (17), the infrared limit is given by :

$$N_b(k \sim 0) \approx \frac{k_B T}{\hbar c_s k}\frac{W_b}{(W_b + \beta k_B T/\hbar)} \sim \frac{1}{k} . \tag{26}$$

This dependence at small *k* is similar to the equilibrium expression. If one now considers the diffusive dispersion of Eq. (20), we find $\mathrm{Re}(\omega_k) = 0$ and $\mathrm{Im}(\omega_k) \sim k^2$, which leads to $N_b(k \sim 0) \sim W_b / k^2$, therefore differing from the equilibrium case. Nevertheless, the relevant physical quantity to the computation of both the condensate depletion and the first order coherence is the state occupation $\langle \psi_k \psi_k^* \rangle$, which is related with the bogolon number as

$$\langle \psi_k \psi_k^* \rangle = N_b(k)\left(|A_k|^2 + |B_k|^2\right) \approx \frac{m k_B T}{\hbar^2 k^2}\frac{W_b}{(W_b + \beta k_B T/\hbar)} \sim \frac{1}{k^2} . \tag{27}$$

Again, the form of the polariton occupation is similar to the equilibrium expression, with a correction due to the finite lifetime of polaritons. The shape of the long-range first order coherence as computed from (28) is therefore expected to be similar to the one obtained for equilibrium systems, but with a quantitative correction induced by the lifetime. On the other hand, the faster decay at large wave-vectors is expected to reduce the short-range decay of the first order coherence. If, in turn, one considers the diffusive dispersion (20), the quantity $|A_k|^2+|B_k|^2$ is constant at low wave vector, which, combined with the bogolon occupation, yields a state occupation $\langle\psi_k\psi_k^*\rangle$ proportional to $1/k^2$, similar to both our model and the equilibrium situation, as pointed out in [31]. The comparison between the two models is given in Table 1. One can see that the final result concerning the long-range coherence of the condensate remains unchanged, being identical to the thermal equilibrium case, at least in the framework of the Bogoliubov approximation.

In order to estimate the first-order coherence of a quasi-condensate state, we rewrite the ansatz (14) using the Madelung transformation, $\psi(\vec{r},t)=\sqrt{n(\vec{r},t)}e^{iS(\vec{r},t)}$. Defining the bogolon creation and annihilation operators, the phase operator can be written as (we omit the time dependence, for convenience)

$$S(r)=\frac{1}{\sqrt{2S}}\sum_{k\neq 0} i\left(\frac{mc_s}{\hbar n_c k}\right)^{1/2}\left(b_k e^{ikr}-b_k^+e^{-ikr}\right). \tag{28}$$

The first order coherence is proportional to $e^{-(\chi(0)-\chi(r))}$, where $\chi(r)=\langle S(0)S^+(r)\rangle$. For one-dimensional systems, we have

$$\chi(0)-\chi(r)\approx\int_0^\infty \frac{mc_s}{\hbar n_c k}N_b(k)\left(1-\cos(kr)\right)dk\approx\frac{W_b}{W_b+\beta k_B T/\hbar}\frac{mk_BT}{2n_c\hbar^2}r. \tag{29}$$

As a result, the first order coherence decays exponentially at large distances, just as in the equilibrium case,

$$g_1(r)\sim\exp\left(-\frac{r}{r_0}\right), \tag{30}$$

where $r_0=\frac{2n_c\hbar^2}{mk_BT}\left(1+\frac{\hbar W_b}{\beta k_B T}\right)$ is the coherence length. The role played by the finite lifetime is to reduce the bogolon occupation with respect to the equilibrium case, which leads to a larger coherence length. As one can see, $r_0$ is inversely proportional to the mass of the particles. For polaritons, with $m$=5 $10^{-5}$ $m_0$ ($m_0$ is the free electron mass), T=5 K and a typical condensate density $n_c$=$10^{10}$ cm$^{-2}$ in a 3 µm wide wire, one finds $r_0>2$ mm, which is one order of magnitude larger than the size of the typical samples studied experimentally [35]. At higher temperature and lower density, $r_0$ can drop to less than 100 µm and become observable experimentally. Similarly, the 2D first-order coherence remains similar to the equilibrium one and reads

$$g_1^{2D}(r) \sim \left(\frac{r_0^{2D}}{r}\right)^{\nu}, \tag{31}$$

with $r_0^{2D} = \frac{\hbar c_s}{k_B T}, \nu = \frac{W_b}{W_b + \beta k_B T/\hbar} \frac{m k_B T}{2\pi\hbar^2 n_c^{2D}}$. For the same typical polariton parameters, we estimate the coherence parameter $r_0^{2D}$ to be of the order of 1.5 µm and $\nu \approx 5\text{x}10^{-4}$, which in practice yields an extremely slow decay of the coherence [47]. We notice that other mechanisms, although neglected in the present model, such as the disorder [48, 49, 50], the finite temporal coherence, and both temporal and spatial fluctuations of the pumping, can affect the spatial coherence of a condensate. However, as far as polaritons are concerned, the standard mechanism of phase fluctuations is highly suppressed because of the low particle mass, and this suppression is further enhanced by the finite particle lifetime.

### *2. Disorder effect on the condensate at rest*

In the presence of disorder, the scattering rates entering the Eqs. (21-23) are given by

$$W_{b,k}^{in} = W_{b,k}^{out} = w_{dis}\delta\left(\text{Re}(\hbar\omega_k) - \hbar\omega_c\right), \tag{32}$$

which in turn yields

$$N_b(k) = \frac{w_{dis}}{-\text{Im}(\omega_k)}\delta\left(\text{Re}(\hbar\omega_k) - \hbar\omega_c\right). \tag{33}$$

The density of bogolons created due to the disorder, calculated within the hybrid BGP model, is strictly zero. There is no disorder-induced depletion of the condensate which is consistent with a dispersion which verifies the Landau criterion. The wave vector dependence of the bogolon occupation given by the diffusive model scales as $1/k^2$, which is similar with the thermal distribution, and the effects on coherence are therefore the same.

## V. Condensate put into motion

When a condensate is put into motion, it is not in the lowest kinetic energy state. The superfluid property however implies that such a motion is stable against the creation of the elementary excitations. Different types of instabilities can be identified. The parametric process which leads to the creation of a bogolon can be resonant with the condensate. In such a case, the imaginary part of a bogolon state becomes positive and the excitation grows exponentially, leading to a dynamical instability of the initial condensed state. The bogolon energy can also be negative (which in general arises for a flow velocity above $c_s$). In the latter case, the condensate is energetically unstable against the growth of low energy bogolons. Finally, in the presence of disorder, the flow decays by elastic scattering, if the bogolon states are available for such scattering at the energy of the condensate. A system in which the dispersion of the excitations verifies the Landau criterion is stable against the three above-mentioned mechanisms. In this section, we investigate the condensate stability against these different processes. We will consider a condensate propagating with a velocity $v_c$ corresponding to a wave vector $k_c$.

### *1. Hybrid BGP model*

The condition (10) states that the gain compensates the life-time at the ground state, which allows the macroscopic occupation of the ground state. For a condensate in motion, the critical conditions are slightly different and read

$$\left(W_{in}^{k_c} - W_{out}^{k_c}\right) - \gamma_{k_c} = \kappa - \beta \frac{\hbar^2 k_c^2}{2m} = 0 \ , \qquad \beta > 0 \tag{34}$$

which yields slightly modified pumping conditions for condensation. The damped Gross-Pitaevskii equation should be rewritten as

$$i\hbar \frac{\partial \psi_k}{\partial t} = \frac{\hbar^2 k^2}{2m}\psi_k + \alpha \int \left(|\psi|^2 + n_R\right)\psi e^{i\vec{k}\vec{r}}\, d\vec{r} - i\hbar\beta \left( \vec{k}_r.\vec{v}_c + \sqrt{\left(\frac{\hbar^2 k_r^2}{2m}\right)^2 + 2\alpha n_c \frac{\hbar^2 k_r^2}{2m}} \right)\psi_k , \tag{35}$$

where $\vec{k}_r = \vec{k} - \vec{k}_c$. The dispersion of the elementary excitations experiences a Galilean boost and is given by

$$\hbar\omega_k = (1 - i\beta)\left( \hbar\vec{k}_r.\vec{v}_c + \sqrt{\left(\frac{\hbar^2 k_r^2}{2m}\right)^2 + 2\alpha n_c \frac{\hbar^2 k_r^2}{2m}} \right). \tag{36}$$

The dispersion (37) is depicted in Figs. 4-a and 4-d. The Landau criterion is verified if $v_c < c_s$ . The bogolon occupation as given by Eq. (33) is zero. Therefore, in this velocity range, the flow is both dynamically and energetically stable. It shows no dissipation against weak disorder and is therefore superfluid. In the supersonic regime, both the real and imaginary part of the bogolon energy change sign with respect to the subsonic regime (see Fig. 4-d). The condensate becomes both dynamically and energetically unstable. This is a difference with the equilibrium situation, when the transition towards the supersonic regime is only accompanied by an energetic instability.

### *2. Diffusive Ginzburg-Landau Model*

We consider Eq. (1-b), for which the dispersion of elementary excitation for a flowing condensate is given by

$$\hbar\omega_{\vec{k}} = \hbar\vec{k}_r.\vec{v}_c - i\Lambda\alpha n_c \pm \sqrt{\left(\frac{\hbar^2 k_r^2}{2m}\right)^2 + \alpha n_c \left(\frac{\hbar^2 k_r^2}{m}\right) - \alpha^2 n_c^2 \Lambda^2} \ . \tag{37}$$

The features of (37) are illustrated in Figs. 4-c and 4-f. In the range $\vec{k}_r.\vec{v}_c < 0$, $\mathrm{Im}(\omega_k) \sim k^2, \mathrm{Re}(\omega_k) \sim -k$. The quantity $W_{b,k}^{in} - W_{b,k}^{out} + \mathrm{Im}(\omega_k) = \mathrm{Im}(\omega_k) - \frac{W_b}{k_B T}\mathrm{Re}(\omega_k) \sim k$ is positive, which corresponds to an amplification of the bogolon occupation by thermal processes. The condensate is therefore energetically unstable both in the subsonic and supersonic regime. A condensate described by the diffusive Ginsburg-Landau model is not superfluid in the Landau sense, both at zero and at finite temperature.

### *3. Diffusive Pitaevskii Model*

We consider the model described in Eq. (1-c). Once critical conditions allowing condensation in a state with finite velocity are fulfilled, the bogolon dispersion reads:

$$\hbar\omega_{\vec{k}} = \hbar\vec{k}_r.\vec{v}_c - i\Lambda\left(\alpha n_c + \frac{\hbar^2 k_r^2}{2m}\right) \pm \sqrt{\left(\frac{\hbar^2 k_r^2}{2m}\right)^2 + \alpha n_c\left(\frac{\hbar^2 k_r^2}{m}\right) - \alpha^2 n_c^2 \Lambda^2 - \hbar\vec{k}_r.\vec{v}_c\left(2i\Lambda\left(\frac{\hbar^2 k_r^2}{2m} + \alpha n_c\right) + \left(\hbar\vec{k}_r.\vec{v}_c\right)\right)}$$

(38)

In Figs. 4-b) and 4-e), we show the real and imaginary parts of (38). At low k, the solution with the plus sign reads

$$\hbar\omega_{\vec{k}} \approx -i\frac{\hbar^2 k_r^2}{2m}\left(\Lambda + \frac{m\left(c^2 - v_c^2\right)}{\alpha n_c \Lambda}\right) - \frac{\Lambda\hbar^3 k_r^2 \vec{k}_r.\vec{v}_c}{m}. \qquad (39)$$

The real part of the energy features a $k^3$ dependence. The imaginary part keeps its negative sign and $k^2$ dependence. The condensate propagating with a subsonic velocity is therefore dynamically and energetically stable. It becomes dynamically unstable in the supersonic regime because of the change of sign of the imaginary part of the energy.

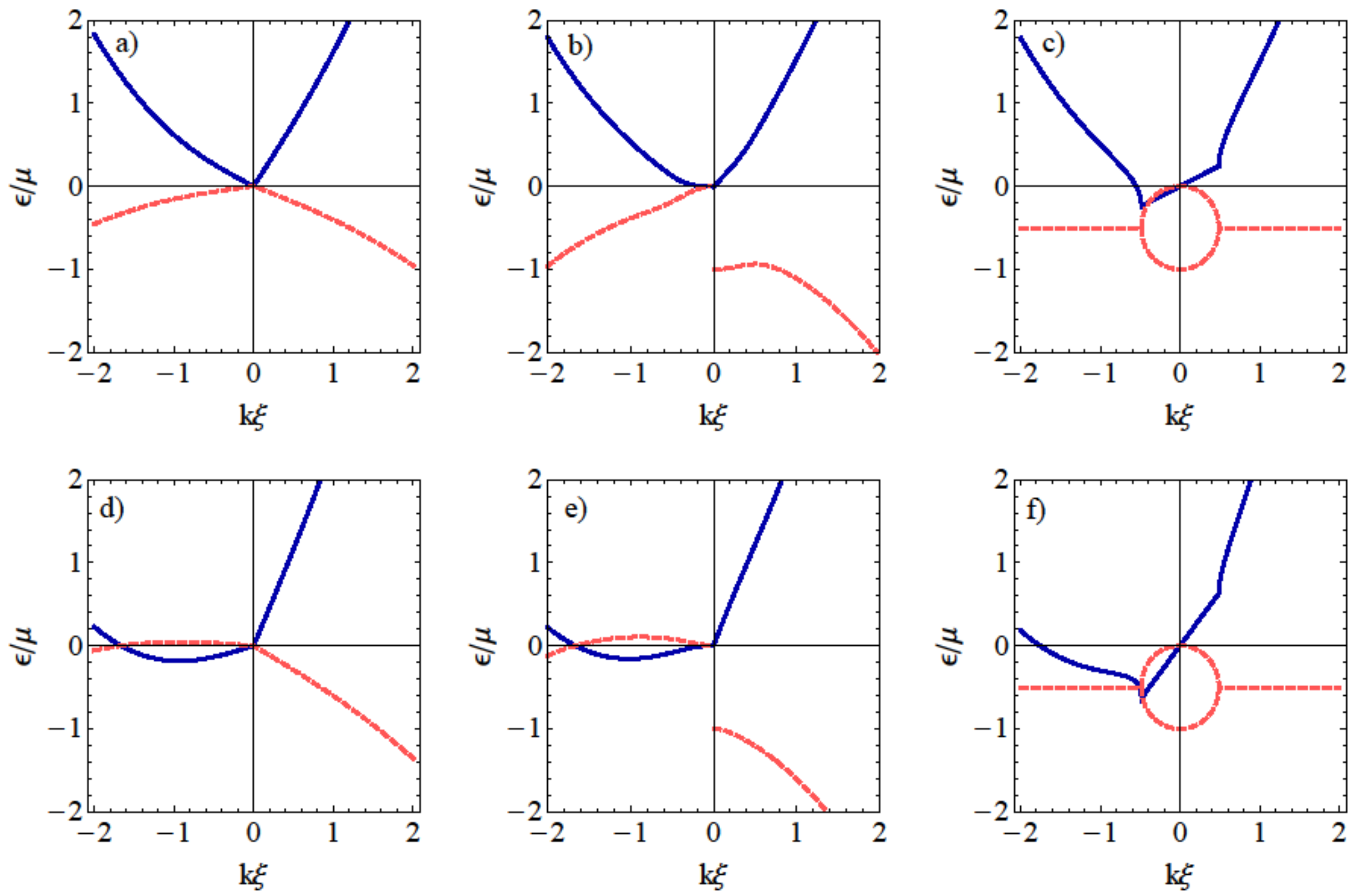

Fig. 4. (color online) Dispersion relation of the elementary excitations for a condensate moving with velocity $v_c = \hbar k_c / m$, illustrated for the three models (hybrid BGP in panels a) and d) with $\beta = 0.25$, diffusive Pitaevskii in panels b) and e) for $\Lambda = 0.5$, and diffusive Ginzburg-Landau for $\Lambda = 0.5$ in c, f). The top panels a), b) and c) depict the subsonic regime $v_c = 0.5 c_s$, while the bottom panels d), e) and f) are obtained for the supersonic regime $v_c = 1.2 c_s$.

## VI. Scattering of the flow against a distribution of delta barriers

In this last section, we analyze the condensate propagation in the presence of disorder, which for convenience is represented by a random distribution of delta barriers, $V_0 \sum_i \delta(\vec{r} - \vec{r}_i)$. The flow is described by a plane wave $\frac{1}{L^{d/2}} e^{i(\vec{k}_c \vec{r} - \omega_c t)}$. The disorder potential scatters condensed particle towards the bogolon states, which results in a mechanical friction. The Boltzmann equation describing particle exchange between the condensate and bogolon states reads

$$\frac{dN_c}{dt} = \sum_k -\left(W_{b,k}^{in} - W_{b,k}^{out}\right) N_b(k) - W_{b,k}^{in}, \tag{40}$$

which, using (32) yields

$$\frac{dN_c}{dt} = -w_{dis} DOS^d(\omega_c) \tag{41}$$

with $DOS^d(\omega_c)$ being the density of bogolon states in a system of dimensionality $d$ at the energy of the condensate. Eq. (41) demonstrates a net decay of the condensate density. In a closed system, this decay - taking place in the supersonic regime - leads to a complete breaking of the flow. The situation is different in an open system. If the decay rate is finite, it is just another source of decay contributing to the radiative decay time, which can be compensated by increasing pumping. Therefore, the steady-state flows and vortex states can be maintained even if some mechanical friction is present. On the other hand, a divergent density of states - associated to a dynamical instability - leads to an infinite decay rate which cannot be compensated by pumping. The flow is therefore unstable against disorder in this case. The decay rate of the condensate given by Eq. (41) is similar to the one resulting from Fermi's Golden rule, which is standardly used to calculate the drag force acting on the condensate [51]

$$\Gamma_{dis}^{d} = \frac{2\pi}{\hbar}\sum_{\vec{k}} \left|M_{\vec{k}}\right|^2 \delta\left(\omega_k - \omega_c\right), \tag{42}$$

where $\left|M_{\vec{k}}\right|^2 = \frac{n_{defect}}{L^d}\left|V_0\right|^2$ and $n_{defect}$ is the density of defects. Eq. (42) can be rewritten as

$$\Gamma_{dis}^{d} = \frac{n_{defect}\left|V_0\right|^2}{\left(2\pi\right)^{d-1}\hbar} DOS^d(\omega_c), \tag{43}$$

which provides an explicit expression for $w_{dis}$,

$$w_{dis} = \frac{n_{defect}\left|V_0\right|^2}{\left(2\pi\right)^{d-1}\hbar}. \tag{44}$$

The drag force induced by the potential on the flow can be written as [51]

$$F^d = \Gamma_{dis}^{d}\frac{\hbar\omega_c}{v}. \tag{45}$$

***1. Hybrid BGP model***

The calculation of the decay rate induced by the potential, and of the corresponding drag force is well known for the Bogoliubov dispersion associated with the Hybrid BGP model:

$$\begin{cases} v < c_s,\ \Gamma_{dis}^{2D} = 0,\ F_{dis}^{2D} = 0 \\ v > c_s,\ \Gamma_{dis}^{2D} \approx \frac{m n_{defect}\left|V_0\right|^2}{\hbar^2}\left(1 - \frac{v^2}{c^2}\right),\ F_{dis}^{2D} \approx \frac{m^2 n_{defect}\left|V_0\right|^2}{2\hbar^2}\frac{v^2 - c^2}{v} \end{cases} \tag{46}$$

$$\begin{cases} v < c_s,\ \Gamma_{dis}^{1D} = 0,\ F_{dis}^{1D} = 0 \\ v > c_s,\ \Gamma_{dis}^{1D} = \frac{n_{defect}\left|V_0\right|^2}{v\hbar^2},\ F_{dis}^{1D} = \frac{m n_{defect}\left|V_0\right|^2}{4\pi\hbar} \end{cases} \tag{47}$$

In Fig. 5 (black solid line), we plot the drag force in 1D and 2D resulting from the analytical calculation of Eq. (45), being in agreement with the approximate results of Eqs. (46) and (47)

### *2. Diffusive Ginzburg-Landau model*

The drag force can be computed semi-analytically, since a numerical integration over the polar angle $\theta$ (arising from the scalar product $\vec{v}_c.\vec{k}_r = v_c k_r \cos\theta$) is necessary. We plot the force associated with the dispersion (37) in Fig. 5 (red/gray solid line), showing the drag force in both 1D and 2D as already performed in [31] and [18]. As it was previously noticed, the drag force shows a clear threshold behaviour at $v_c = c_s$ .

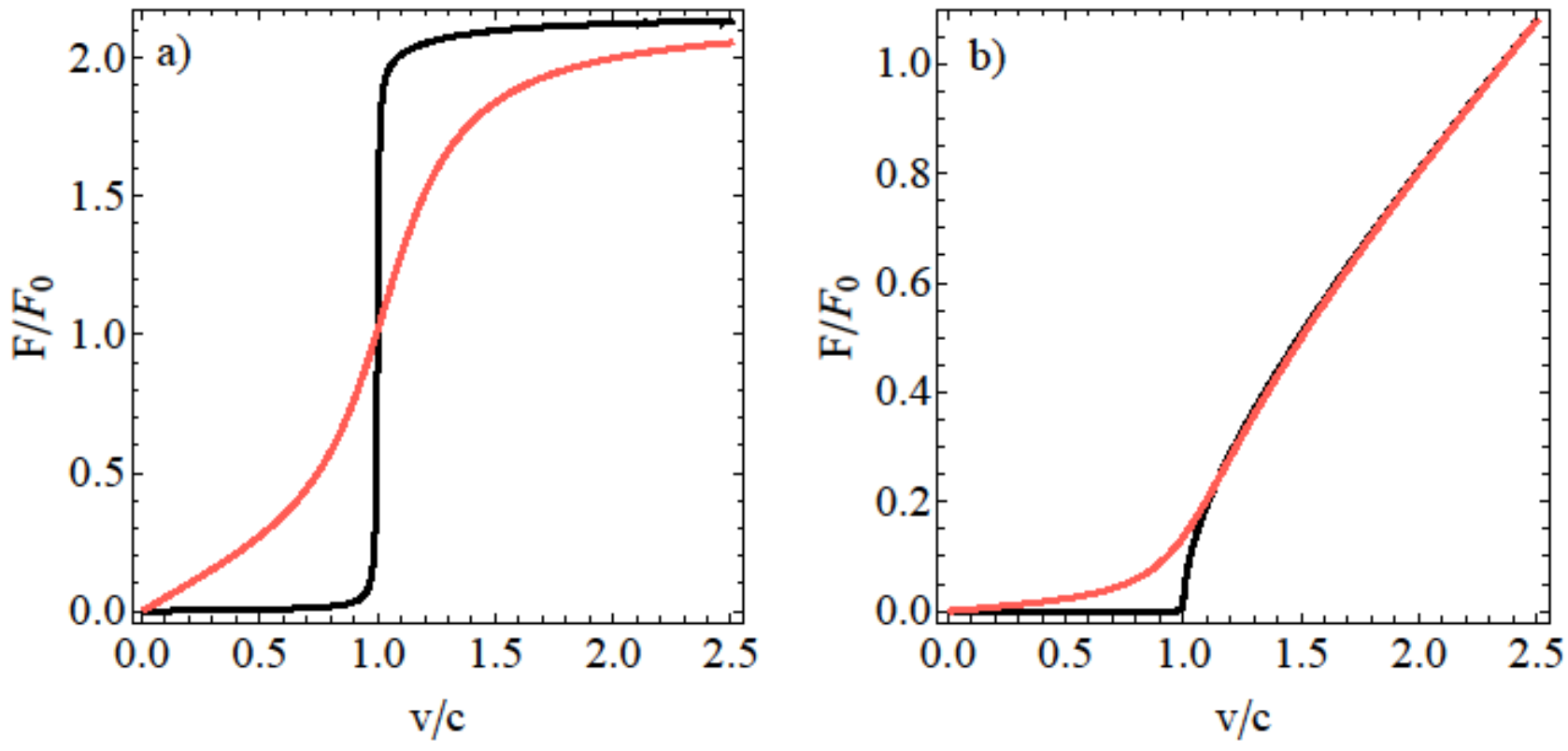

Fig. 5. (color online) Drag forces experienced by an infinite flowing condensate past a density of obstacles, for 1D (panel a), in units of $F_0 = mn_{defect}|V_0|^2/(8\pi\hbar)$ ) and 2D (panel b), in units of $F_0 = m^2 n_{defect}|V_0|^2/(2\hbar^2)$) systems. In both situations, the black curve corresponds to hybrid Boltzmann-Gross-Pitaevskii model with, while the red/gray line illustrates the case of the diffusive Ginzburg-Landau model with $\Lambda = 0.65$.

## VII. Discussion and conclusion

The main features of the conservative, hybrid BGP and diffusive models are summarized in Table 1. With respect to the conservative case, the hybrid BGP model shows several differences. Condensation occurs only above some minimal temperature, ensuring a fast enough relaxation kinetics compared with the particle lifetime. The first-order coherence shows similar qualitative features in both cases, whereas the coherence length is longer in the hybrid BGP model due to the finite bogolon lifetime. When the condensate is put into motion, both hybrid BGP and the two diffusive models show a stable superfluid motion below the sound speed. In the supersonic regime, however, the hybrid BGP model develops energetic and dynamical instabilities, which departs from the conservative case which is only energetically unstable. The diffusive Pitaevskii model does not describe a stable condensate in motion, since its dispersion scales as $k^3$ for low wavectors, in the subsonic regime, and becomes both energetically and dynamically unstable in the supersonic regime. Dynamical instability, on the other hand, is not contained in the diffusive Ginzburg-Landau model, but the latter clearly violates Landau's criterion of superfluidity. According to Ref. [18], superfluidity in non-equilibrium systems can still be recovered since the excitations due to a defect do not propagate. This local perturbation of the fluid in the presence of a defect has been pointed as one of the possible signatures of superfluidity in driven open-dissipative systems. Our model, however, accounts for the possibility of a driven system to satisfy the usual Landau's criterion of superfluidity.

| | Conservative | Open Hybrid BGP | Open Diffusive Ginzburg-Landau | Open Diffusive Pitaevskii |
|---|---|---|---|---|
| Temperature range for quasi-condensation | $T < T_c^d$ | $T_{c_1} < T < T_{c_2}$ | Not Applicable | Not Applicable |
| $v_c = 0$ $N_b(k)$ | $1/k$ | $1/k$ | $1/k^2$ | $1/k^2$ |
| $v_c = 0$ $\langle \psi_k \psi_k^* \rangle$ | $1/k^2$ | $1/k^2$ | $1/k^2$ | $1/k^2$ |
| $v_c = 0$. $g_1^{1D}(r)$ | $\exp\left(-\frac{r}{r_0}\right)$ $r_0 = \frac{2n_c\hbar^2}{mk_BT}$ | $\exp\left(-\frac{r}{r_0'}\right)$ $r_0' = r_0\left(1+\frac{\hbar W_b}{\beta k_B T}\right)$ | $\exp\left(-\frac{r}{r_0''}\right)$ $r_0'' = \frac{n_c\hbar^2}{mW_b\hbar\Lambda}$ | $\exp\left(-\frac{r}{r_0''}\right)$ $r_0'' = \frac{n_c\hbar^2}{mW_b\hbar\Lambda}$ |
| $v_c = 0$ $g_1^{2D}(r)$ | $\left(\frac{r_0^{2D}}{r}\right)^{\nu}$ $r_0^{2D} = \frac{\hbar c_s}{k_B T}$, $\nu = \frac{mk_BT}{2\pi\hbar^2 n_c^{2D}}$ | $\left(\frac{r_0^{2D}}{r}\right)^{\nu'}$ $\nu' = \nu\frac{W_b}{W_b + \beta k_B T/\hbar}$ | $\left(\frac{r_0^{2D}{}'}{r}\right)^{\nu''}$ $r_0^{2D'} = \frac{\hbar c_s}{2mW_b\hbar\Lambda}$, $\nu = \frac{mW_b\hbar\Lambda}{\pi\hbar^2 n_c^{2D}}$ | $\left(\frac{r_0^{2D}{}'}{r}\right)^{\nu''}$ $r_0^{2D'} = \frac{\hbar c_s}{2mW_b\hbar\Lambda}$, $\nu = \frac{mW_b\hbar\Lambda}{\pi\hbar^2 n_c^{2D}}$ |
| $v_c < c_s$ Dynam. Stab. | Yes | Yes | Yes | Yes |
| $v_c > c_s$ Dynam. Stab | Yes | No | Yes | No |
| $v_c < c_s$ Energ. Stab. | Yes | Yes | No | Yes |
| $v_c > c_s$ Energ. Stab. | No | No | No | No |
| $v_c < c_s$ Superfluid | Yes | Yes | No | No |
| $v_c > c_s$ Superfluid | No | No | No | No |

Table 1: Summary of the main condensate features described by the conservative, Hybrid BGP and diffusive models.

In conclusion, we introduced a closed set of equations to describe the relaxation of a homogeneous BEC coupled to a reservoir at a finite temperature. Our model, referred to as the hybrid BGP model, consists of a Gross-Pitaevskii equation for the condensate and a semi-classical Boltzmann equation describing a pumped, spatially homogeneous two-dimensional excitonic reservoir. A finite energy gap is assumed between the exciton reservoir and the polariton condensate. We established the critical condition needed for condensation in the polariton ground state. If such a critical condition is satisfied, we find that the GP equation describing the elementary excitation contains a dissipative term proportional to the energy of the elementary excitations of the system, and not to the total energy including its non-linear interaction term, as considered previously in [6,7,18,30]. As a result, the real part of the energy of the elementary excitations (bogolons) satisfies the Landau criterion of superfluidity, whereas the imaginary part corresponds to the relaxation. It supports a superfluid stable flow. The decay of the first-order spatial coherence at finite temperature, induced by phase fluctuations, takes a similar form as in the case of closed systems, with an exponential decay in 1D and power-law decay in 2D. We found that the coherence lengths in the presence of decay processes are larger than their equilibrium counterparts. The low polariton mass makes the phase fluctuation effects in general extremely weak. We find that the results of the equilibrium models are relevant as a limiting case of the dissipative model. Although we focus on the description of exciton-polaritons condensates, our approach may be an important contribution to the understanding of BEC dynamics in other systems, which are also only approximately conservative, providing a simple, yet systematic approach to the phenomenology of dissipation in driven condensates.

***Acknowledgments***. We would like to thank Jacqueline Bloch and Alberto Amo for the stimulating discussions we had on this topic. We acknowledge the support of the Labex GANEX, and of the ANR Quandyde, and IRSES Polaphen (grant No 246912).

---